\documentclass[letterpaper,journal]{IEEEtran}
\usepackage{amsmath,amsfonts}
\usepackage{algorithmic}
\usepackage{algorithm}
\usepackage{array}
\usepackage[caption=false,font=normalsize,labelfont=sf,textfont=sf]{subfig}
\usepackage{textcomp}
\usepackage{stfloats}
\usepackage{url}
\usepackage{verbatim}
\usepackage{graphicx}
\usepackage{cite}
\usepackage{xcolor}
\usepackage{soul}
\usepackage{cite}
\usepackage{hyperref}
\begin{document}


\title{Detector Asymmetry in Continuous Variable Quantum Key Distribution }{}

\author{Jennifer O Bartlett, Alfie J Myers Wilson, Christopher J Chunnilall and  Rupesh Kumar
\thanks{This paper was produced by the IEEE Publication Technology Group. They are in Piscataway, NJ.}
\thanks{Manuscript received June 1, 2025; revised August 1, 2025.}}



\maketitle
  
\begin{abstract}
In Local-local Oscillator (LLO) based Continuous-Variable Quantum Key Distribution (CV-QKD), the phase reference of the transmitter and receiver, Alice and Bob, are naturally de-correlated due to their use of individual lasers. A phase reference signal is used, whose measurement is critical for estimating the phase difference and correcting the raw QKD data. We observed that asymmetry in the quadrature measurements of the shot noise-limited heterodyne detector affects the accuracy of the reference signal's phase estimation and thereby reduces the achievable transmission distance and key rate of the CV-QKD system. We quantify the effect and propose a method to counteract the effect of detection asymmetry. We also evaluate the effects of detection asymmetry using quantum optical tomography.

\end{abstract}


\section{Introduction}
\IEEEPARstart{I}{n} the last few decades, advances in technology have been shown to threaten the security of classical communication \cite{rusca2024quantum}. Quantum Key Distribution (QKD) is a solution that offers information-theoretic security based on the laws of physics, facilitating the secure distribution of encryption keys between two authenticated users, Alice and Bob\cite{BENNETT20147}. 

In Continuous Variable Quantum Key Distribution (CV-QKD), information is encoded onto the continuous properties of the electromagnetic field, such as amplitude and phase \cite{LITREVIEW_Laudenbach_2018}. It shares similar architectures with coherent optical communications, except that the signals are at the quantum level and information decoding relies on shot-noise-limited detection. The Heisenberg uncertainty principle provides security against information eavesdropping. One of the main benefits is that CV-QKD integrates with the classical communication network to provide cost-effective room temperature technology\cite{diamanti2015distributing}. Furthermore, the information can be encoded in higher-dimensional spaces, allowing higher key rates at short distances \cite{grosshans2003quantum}. 
At longer  distances,  key generation becomes challenging due to greater excess noise. However, recent demonstrations have shown successful key generation over 202.81 km and 100 km for the two main CV-QKD schemes \cite{zhang2020long, hajomer2024long} . For security purposes, it is assumed that all excess noise originates from eavesdropping activities, although in practice, non-ideal state preparation and measurement can contribute \cite{RUIZCHAMORRO2023e16670, silva2020practical}.

To perform CV-QKD, Alice modulates the amplitude and phase of her attenuated source to generate quadrature-modulated quantum signals, most commonly following the GMCS (GG02) protocol \cite{Grosshans2002}. She sends these to Bob over a public quantum channel using her preferred implementation. Bob extracts the quantum information by interfering the attenuated quantum signal with a strong reference signal, the Local Oscillator (LO). This is performed by homodyne or heterodyne detection \cite{ grosshans2003quantum, weedbrook2004quantum}, where the LO acts as a calibration mechanism between Alice and Bob, setting the reference frame for the measurement of quantum signals. The LO also performs mode filtering by its interference with a quantum signal of matching coherence and reduces the impact of noise photons from the quantum channel during Bob's measurement \cite{qi2010feasibility}.  This makes CV-QKD an attractive candidate for highly populated fibre optic networks and free-space channels \cite {kumar2015coexistence, hajomer2025coexistence, rani2023freespacecontinuousvariable}. 

One of the most common implementations of CV-QKD is using the Transmitted Local Oscillator (TLO), in which the LO is co-transmitted alongside the quantum signal \cite{lodewyck2007quantum}. Despite this,  LO leakage/crosstalk \cite{ qi2007experimental}, LO launch power, and LO attacks \cite{Ma2013, jouguet2013preventing, huang2013quantum, huang2014quantum}, all constrain this approach. This is because the LO serves not only as a calibration method, but also as a reference for the shot-noise measurement and clock generation. Moreover, the transmitted LO signal intensity can be manipulated, affecting the shot-noise variance, and compromising the detection sensitivity of the homodyne/heterodyne detectors at Bob \cite{chi2011balanced}. 

Modifications have since been devised to use a Local-local oscillator (LLO), where two independent but highly coherent lasers generate and detect the quantum signals locally at Alice and Bob, respectively \cite{Huang:15, Qi2015}. In this scenario, an eavesdropper, Eve, can no longer access the LO, limiting the types of attacks she can perform on the transmission. This also means that Alice and Bob now have two independent free-running laser sources that create two misaligned frame of reference \cite{Soh2015}.  As a result, Bob's measurement outcomes are inherently uncorrelated to the quantum signal generated by Alice. Therefore, the alignment of the reference frame is a distinctive problem, where the information defining the reference frame cannot be conveyed through conventional means. To address this, many pilot-aided schemes have been developed\cite{kleis2017continuous, Marie_2017, Wang2018, Shao2022,Wang:23, hajomer2024long}.

The common technique is to send a phase reference signal alongside the quantum signal.  Bob measures the reference signals with a heterodyne detector; a pair of shot noise-limited homodyne detectors, each measuring their respective conjugate quadrature of the input signal. From the measurement outcome, Bob estimates the reference phase and either corrects his measurement outcome or discloses the reference phase over the public channel for Alice to post-correct her data.  Accurate measurement of the phase of the reference signal determines the degree of correlation between Alice and Bob. The excess noise due to the reference signal's phase estimation error is a limiting factor for achieving long-distance key distribution in the LLO scheme.

In this paper, we consider heterodyne detection for reference phase estimation and study the impact of detection asymmetry between signal quadrature measurements on the excess phase noise. We propose a scaling method to correct the detection asymmetry and reduce the effect of reference frame misalignment on the transmission distance.  We also considered the impact of heterodyne asymmetry in quantum optical state tomography by reconstructing their respective Wigner plots.  We show that estimating the Wigner function from the asymmetric quadrature data shows a reduction in the state's fidelity. 

The paper is structured as follows: Section \ref{sec:level2} details the different types of detector schemes commonly used to measure continuous variables, alongside a description of a typical LLO design and the alignment of the reference frame. Section \ref{sec:level3} is where we identify errors within the detection scheme and show that these errors result from an asymmetry between detectors. In Section \ref{sec:level5}, we present our experimental findings and calibrate the excess noise parameter from this to plot the key rate. Lastly, we conclude our findings in Section   \ref{sec:level6}.

\section{\label{sec:level2}Methods}
\subsection{Detection }

\begin{figure}[!t]
\centering
\includegraphics[width=3.5in]{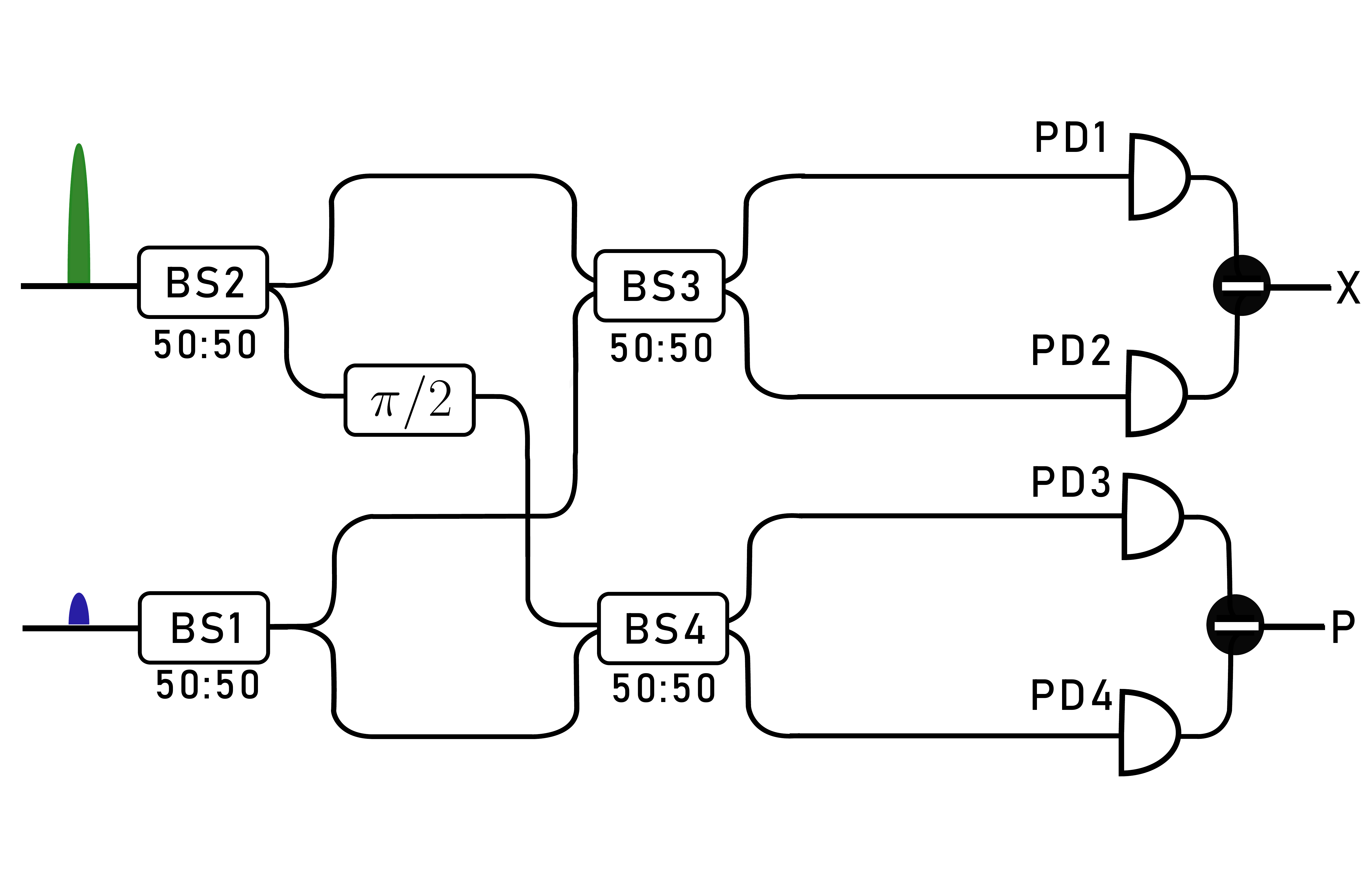}
\caption{A schematic diagram of a heterodyne detector. A $90$ degree phase shift is applied to one branch of the Local Oscillator (LO), allowing full quadrature reconstruction of the $X$ and $P$ phase space. Blue and green pulses indicate the Gaussian modulated quantum signal and bright LO pulse, respectively. }
\label{fig:fig_1}
\end{figure}

In CV-QKD, two primary shot-noise-limited detection schemes are used. In the first method, homodyne detection, the incoming quantum signal interferes with the LO. The phase of the LO is randomly set to $0$ or $\pi/2$ and the respective conjugate quadratures, $X$ and $P$,  are measured.  In the second method, heterodyne detection,  both quadratures are simultaneously measured and, therefore, can be modelled as a dual homodyne detection, see Figure \ref{fig:fig_1}.  In this scheme, the detection uses a 90-degree optical hybrid that splits the incoming quantum signal and LO equally. It then adds the respective phase to each LO branch (0 to one and $\pi/2$ the other) and mixes it with the corresponding signal fields on their respective beamsplitters.  Two balanced photodetectors detect the signals.

Consider the fields incident to the lower branch of the $90$-degree hybrid:
\begin{align}
E_{S} & = \sqrt{\frac{2P_{S}}{k}}  \cos\left( 2 \pi f_{S} t  + \phi_{{S}}(t) \right), \\
E_{LO}& = \sqrt{\frac{2P_{LO}}{k}} \cos\left( 2 \pi f_{LO} t + \phi_{{LO}}(t)  +\varphi_{\pi/2}\right),
\end{align} 

where $P_{S}$ and $P_{LO}$, $f_{S}$ and $f_{LO}$ and $\phi_{\text{S}}(t)$ and $\phi_{LO}(t)$ are respectively the power, centre frequency, and phase of the signal and LO at Bob. $k=D n/{Z_0}$, where $D$ is the effective beam area ($D \approx \pi (\mathrm{MFD}/2)^2$, where $\mathrm{MFD}$ is the mode field diameter in fibre) , $n$ is the refractive index of glass and ${Z_0}$ is the impedance of free space. Channel contributions, such as attenuation, phase rotation, etc., are omitted for ease of derivation.


After $E_{S}$ and $E_{LO}$ are mixed on the beam splitter, $BS4$, the signals are detected by photodiodes $PD3$ and $PD4$. They are converted into photocurrents:

\begin{align}
I_{P+} & = kR_{PD3} \left|\frac{\left( E_{S} + E_{LO}\right)}{\sqrt{2}} \right|^2 , \\
I_{P-} & =kR_{PD4} \left| \frac{\left(E_{S} - E_{LO}\right)}{\sqrt{2}}\right|^2\nonumber,\\
\end{align}
where $R_{PD3(4)}$ is the responsivity of photodiode $PD3(4)$ defined by $R_{PD3(4)}=(q \eta/\hbar \omega)_{PD3(4)}$, where $q$ is the electronic charge, $\eta$ is the detector quantum efficiency and $\omega$ is the frequency of the incident photons \cite{FundamentalsofCVdetection }. We assume that $R_{PD3}=R_{PD4} =R $, as each detector has the same responsivity.  To calculate the final output, both photocurrents are resolved as
\begin{align}
I_{P}&=I_{P+}-I_{P-}\\
&=2R_{} \sqrt{P_{S}P_{LO}} \cdot \cos\left( 2 \pi f_{S} t  + \phi_{{S}}(t) \right) \nonumber\\
&\cdot \cos\left( 2 \pi f_{LO} t + \phi_{{LO}}(t)  +\varphi_{\pi/2}\right),\\
& =   R_{} \sqrt{P_{S}P_{LO}} \cos \left( 2 \pi \Delta f t  + \Delta\phi-\varphi_{\pi/2}\right),
\end{align}

where the sum frequency is too rapid for the detector to resolve and therefore only the difference frequency is resolved \cite{Ingard2008-cy}. The difference frequency is defined as $\Delta \phi=\phi_S-\phi_{LO}$. We can also consider that the LO at Bob is perfectly frequency matched with the signal sent from Alice, therefore $\Delta f=0$. $I_P$ simplifies to
\begin{align}
I_P& = R \sqrt{P_{S}P_{LO}} \cos \left( \Delta\phi -\varphi_{\pi/2}\right),
\label{eq:eq4}
\end{align}

Likewise, the photocurrent output on the upper branch, corresponding to the $X$ quadrature measurement, is 
\begin{align}
I_X & =  R \sqrt{P_{S}P_{LO}} \cos \left( \Delta\phi\right).
\label{eq:eq5}
\end{align}
The magnitude of each photocurrent is therefore mainly determined by the power incident on each detector.  Each photocurrent is amplified and fed into instrumentation for acquisition and data processing.  Since it is assumed that each pair of photodetectors is balanced, the mean value of the current over a large number of measurements is zero.


When using these detection methods, it is important to consider the effects of loss on the incoming quantum signals and how this affects the resulting signal-to-noise ratio (SNR) of the measured signals. Moreover, due to the additional beamsplitters, the heterodyne detector is more lossy than the homodyne detector. Therefore, for practical implementation,  measuring the reference signal with a heterodyne detector and the quantum signal with a homodyne detector is beneficial \cite{Soh2015, Wang:23}. Due to this, and as expanded on in Section \ref{sec:level3}, measurement errors can be imparted by the difference in measurement outcomes between the homodyne detector and heterodyne detector. These are reflected in the final secure key rate in LLO-based CV-QKD schemes. 


\subsection{\label{sec:level3} LLO based CV-QKD scheme}

\begin{figure*}[!t]
  \centering
  \includegraphics[width=\linewidth]{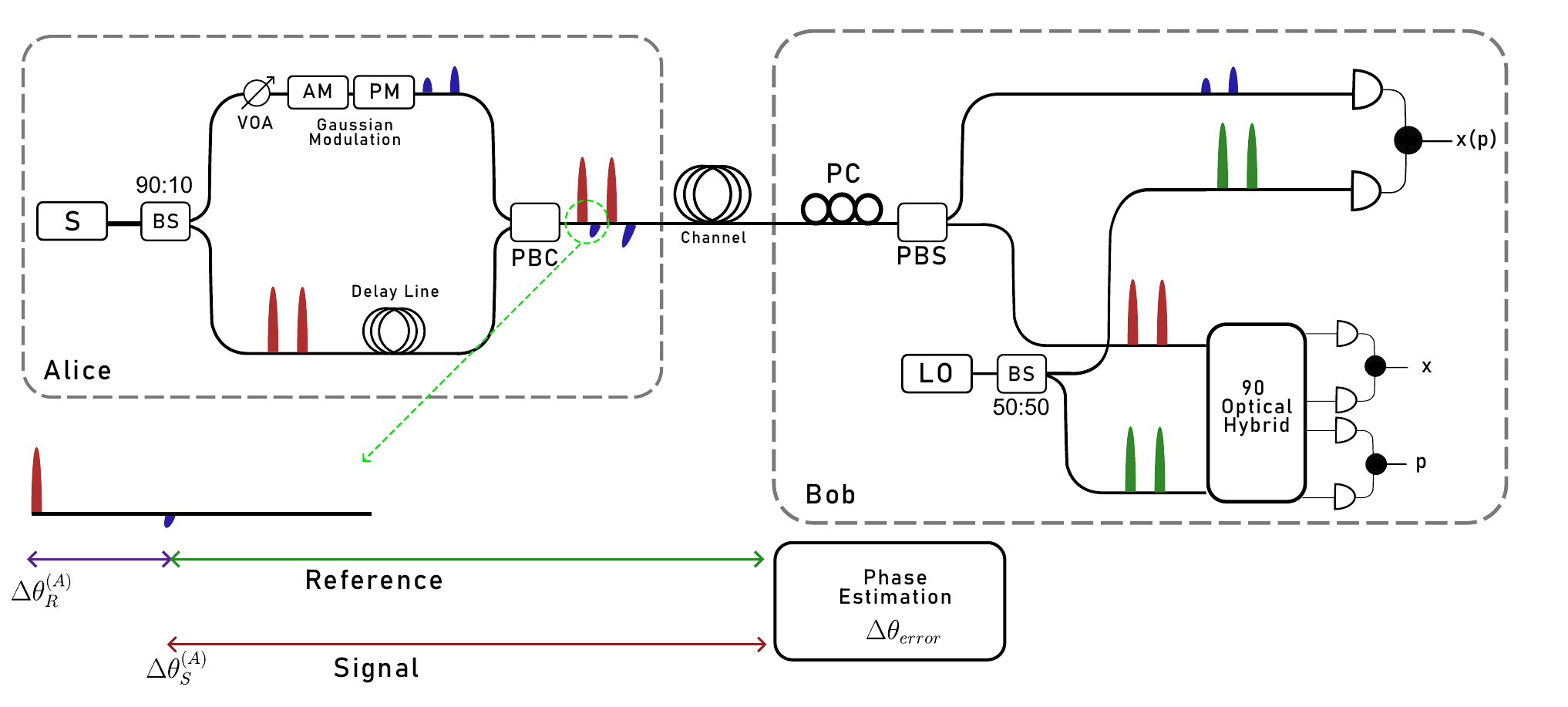}
  \caption{A typical polarisation time-multiplexed LLO design. The bottom-left illustration shows the uncertainty in path position between the reference and quantum signals. \label{fig:LLOdesign}}
\end{figure*}
We consider a version of the time and polarization multiplexed LLO design with pulsed signals as shown in Figure \ref{fig:LLOdesign}, using the detection scheme in \cite{Wang:23}.  The set-up features two distinct  laser sources situated at Alice and Bob. 


Alice carves light  pulses from a continuous wave, highly coherent, narrow linewidth laser and splits this signal with a 90:10 beamsplitter. The strong pulse acts as the reference signal, and the weak pulse is quadrature-modulated and attenuated to near vacuum level. The maximum amplitude of the reference signal is based on the dynamic range of the pulse-carving modulator.  The reference path incorporates a delay line to approximate the path length of the quantum signal branch, which is done to reduce any phase noise associated with path length differences. The reference and signal pulses are multiplexed and sent to Bob over the quantum channel.


Bob uses a polarisation controller (PC) to compensate for polarisation drift from the channel. The signal and reference pulses are demultiplexed. Bob then mixes the reference and quantum signal with his local laser source, the local-local oscillator (LLO), which he can carve as strong pulses or keep as a continuos wave (CW). To increase the SNR and maximise the transmission distance in LLO CV-QKD, one can use homodyne detection for the quantum signal and heterodyne detection for reference signal phase estimation. 

Since Bob uses a locally generated local oscillator for the homodyne/heterodyne detection, reference frame alignment becomes a necessary  procedure for correlating  Alice's  prepared  signal quadrature data with Bob's measured signal quadrature data. This reduces the excess noise induced by  reference frame misalignment  and increases the  secure key generation rate.



\subsection{\label{sec:level4}Reference Frame Alignment}
\label{sec:refframealignment}

In the LLO design, reference frame alignment is essential for making the scheme viable. As discussed in the description in \cite{Marie_2017, Shao2021}, the phase of the quantum signal relative to the LO is given by
\begin{equation}
\Delta \theta_{S}^{(A)}=\phi_{S}^{(A)}-\phi_{LO}^{(A)} \quad \text{and} \quad
\Delta \theta_{S}^{(B)}=\phi_{S}^{(B)}-\phi_{LO}^{(B)},
\label{eq:signal}
\end{equation}
where $\phi_{LO}^{(A(B))}$ is the phase of the LO and $\phi_{S}^{(A(B))}$  is the phase of the quantum signal at Alice(Bob).  Similarly,  the relative phase of  the reference signal with respect to the LO is given  as: 
\begin{equation}
\Delta \theta_{R}^{(A)}=\phi_{R}^{(A)}-\phi_{LO}^{(A)} \quad \text{and} \quad
\Delta \theta_{R}^{(B)}=\phi_{R}^{(B)}-\phi_{LO}^{(B)}.
\label{eq:reference}
\end{equation}
Three fundamental components contribute to the misalignment between the reference frames of Alice and Bob. The first is the inherent phase drift, $V_{drift} $ , between each laser.  This can be defined as:
\begin{equation}
V_{\mathrm{drift}} = 2\pi \cdot \left( \Delta v_A + \Delta v_B\right) \cdot |t_R -t_S|,
\end{equation}
where $\Delta v_A$ and $\Delta v_B$ are the spectral linewidths of the  laser  at Alice and Bob, respectively, and $t_R$ and $t_S$ are the times at  which the reference and signal pulses are created in relation to each other at Alice. 

The second contribution comes from the phase drift between the signal and reference due to their optical path difference. This can be estimated as 
\begin{equation}
V_{\mathrm{path}} \approx \mathrm{var}\left(\Delta \theta_{S}^{(path)} -\Delta \theta_{R}^{(path)}\right),
\end{equation}
where $(\cdot)^{(path)}$ denotes the effects of path differences; however, it is otherwise defined the same as in Equation \ref{eq:signal} and \ref{eq:reference}.

Lastly, there is an estimation error for the phase of the reference signal by Bob.  We find that the actual phase  $\Delta \theta_{R}^{(B)}$  will differ from the estimator $\hat{\Delta \theta}_{R}^{(B)}$. This error can then be approximated as
\begin{equation}
V_{\mathrm{det}} \approx \mathrm{var}\left(\Delta \theta_{R}^{(B)} -\hat{\Delta \theta}_{R}^{(B)}\right), 
\label{eq:goodphaseeq}
\end{equation} which can be shown in the case of no channel contribution as 
\begin{equation}
V_{\mathrm{det}} \approx \mathrm{var}\left(\Delta \theta_{R}^{(B)} -\hat{\Delta \theta}_{R}^{(B)}\right) \equiv \mathrm{var}\left(\Delta \theta_{R}^{(A)} -\hat{\Delta \theta}_{R}^{(B)}\right).
\end{equation} 
We refer to $\Delta \theta_{R}^{(A)}$  as the ``ideal" phase of the reference signal. The phase estimator for the reference signal  is then obtained from  Bob's measurements of each quadrature value  of the reference signal, which is  calculated as
\begin{equation}
\hat{\Delta \theta}_{R}^{(B)}=\tan ^{-1}\left\{\frac{\left<P_R^{(B)}\right>}{\left<X_R^{(B)}\right>}\right\}.
\label{eq:esttheta}
\end{equation}

Since each phase noise contribution is linearly independent of the other, the overall  phase misalignment  can be written as:
\begin{equation}
V_{\mathrm{total}}=\xi_{\mathrm{drift}}+\xi_{\mathrm{path}}+\xi_{\mathrm{det}}.
\end{equation}
The excess noise on the quantum signal from modulation variance  $V_{A}$ , due to reference frame misalignment is
\begin{equation}
\xi_{\mathrm{error}}=V_{\mathrm{A}} \cdot V_{\mathrm{total}}=2V_{\mathrm{A}} \cdot \left(1-e^{-V_{\mathrm{total/2}}} \right),
\end{equation}
\cite{Marie_2017}.
\section{\label{sec:level3}Detector Asymmetry}


As described in Section \ref{sec:level2}, Bob measures the reference signal with a heterodyne detector. After the current subtraction from each pair of photodiodes, the difference in current is proportional to the quadrature amplitudes: $X$ and $P$.  This is deemed an accurate representation of the quadrature values, assuming that each balanced detector is balanced proportionally to the other.   In other words, each pair of photodiodes receives the same amount of light on average across both  $X$ and $P$ measurements.

However,  it is not easy to achieve this in practice.  For example,  for a given input LO power to the $90$-degree hybrid, each pair of balanced detectors gets a different amount of light intensity due to various causes such as:  non-ideal 50/50 splitting ratio of the beam splitters, the imbalance of individual balanced detectors (homodyne imbalance)  \cite{qin2018homodyne, 9203349}, an inaccurate $90$-degree phase shifter \cite{Shen:21}, differences in responsivity between two photodetectors \cite{RUIZCHAMORRO2023e16670} or a difference in coupling loss at the fibre connectors, etc.  As a result, one pair of photodetectors,  for example, those for the $X$ quadrature,  can receive a comparatively higher level of light than those for the $P$ quadrature.  We refer to this as an asymmetry in the heterodyne detection. Nevertheless, the constituent photodiodes' currents can be perfectly balanced, irrespective of the relative difference in  the light received by  the pairs of photodiodes.  For instance, if the variances of the quadrature measurements are measured, one can observe that heterodyne asymmetry leads to different variance levels for each quadrature.  

In such cases, it is essential to scale the inputs of either  measurement to ensure that each pair of photodiodes receives the same amount of light on average.  One solution is to scale the quadrature data using the LO variance measurement; however, noise from this measurement may naturally cause deterioration in the quality of the quadrature data. Therefore, as we do not need to normalise the reference signal into shot noise units (SNU), we can achieve this using attenuators or in post-processing by scaling each measurement such that:
\begin{equation}
X_{\text{scaled}} = \left(\frac{X - X_{\text{min}}}{X_{\text{max}} - X_{\text{min}}}\right) \times (P_{\text{max}} - P_{\text{min}}) + P_{\text{min}}.
\label{eq:normalisation}
\end{equation}
From this, we can create a new set of quadrature data that will correspond to the corrected phase of the reference signal and must be used when correcting the quantum signal phase.


If not accounted for, this error can be attributed to a form of detection noise, $V_{det}$. Since the components of $V_{total}$ are linearly independent, the variance associated with this error can be isolated as 
\begin{equation}
V_{det} \approx \mathrm{var}\left[\mathrm{\left(\Delta \theta_R^{(B)}\right)_{\text{Scaled}}}-\left(\Delta \theta_R ^{(B)}\right)_{\text{Asymmetric}}\right],
\label{eq:gaindiff3333}
\end{equation}
where $\left(\Delta \theta_R^{(B)}\right)_{\text{Scaled}}$ is modelled as the symmetric/ideal detection of the reference signal. Using this, we calculate the phase error's contribution to excess noise as
\begin{equation}
\xi_{\mathrm{det}} =  V_{\mathrm{A}} \cdot  V_{\mathrm{det}}  =  2 V_{\mathrm{A}} \left(1-e^{-V_{\mathrm{det/2}}} \right).
\label{eq:excessNoiseDetector}
\end{equation}


\section{\label{sec:level5}Experimental Setup and Results}

\begin{figure}[!t]
\centering
\includegraphics[width=3.5in]{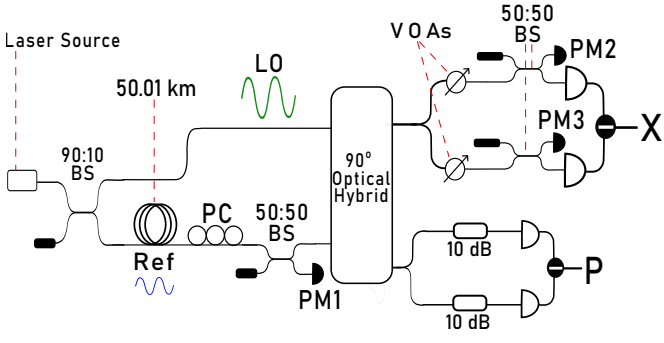}
\caption{The experimental setup shows a 90 degree optical hybrid, where one arm is attenuated by VOA's to demonstrate different asymmetries. When the power is highly attenuated a larger asymmetry is seen.}
\label{fig:experimental_set_up_lab}
\end{figure}

To test the asymmetry of the 90-degree hybrid heterodyne detector, the experimental setup indicated in Figure \ref{fig:experimental_set_up_lab} was used.  Light from a continuous wave (CW), highly coherent laser source (NKT Koheras ADJUSTIK), at 1550 nm, was split by a 90/10 beamsplitter (BS) to produce an LO and reference signal. The reference signal was attenuated  and sent over a 50.01 km fibre spool.   A beamsplitter monitors the reference signal around $E_{R}^2=552$ SNU (55.5 $ V_A$) (at detector P). A  polarization controller  was used to align the polarization with respect to the LO. The reference signal and LO are mixed at a 90-degree hybrid (Optoplex) and sent to the 100 MHz balanced photodetectors (Koheron PD100B).  To mimic various levels of asymmetry, variable attenuators (VOAs) and monitoring beamsplitters change the intensity of input light to one pair of photodiodes for the X quadrature measurement. For the P quadrature measurement, loss-matching fixed attenuators are attached to the other pair photodiodes to keep the light intensity constant.  


\begin{figure}[!t]
\centering
\includegraphics[width=3.5in]{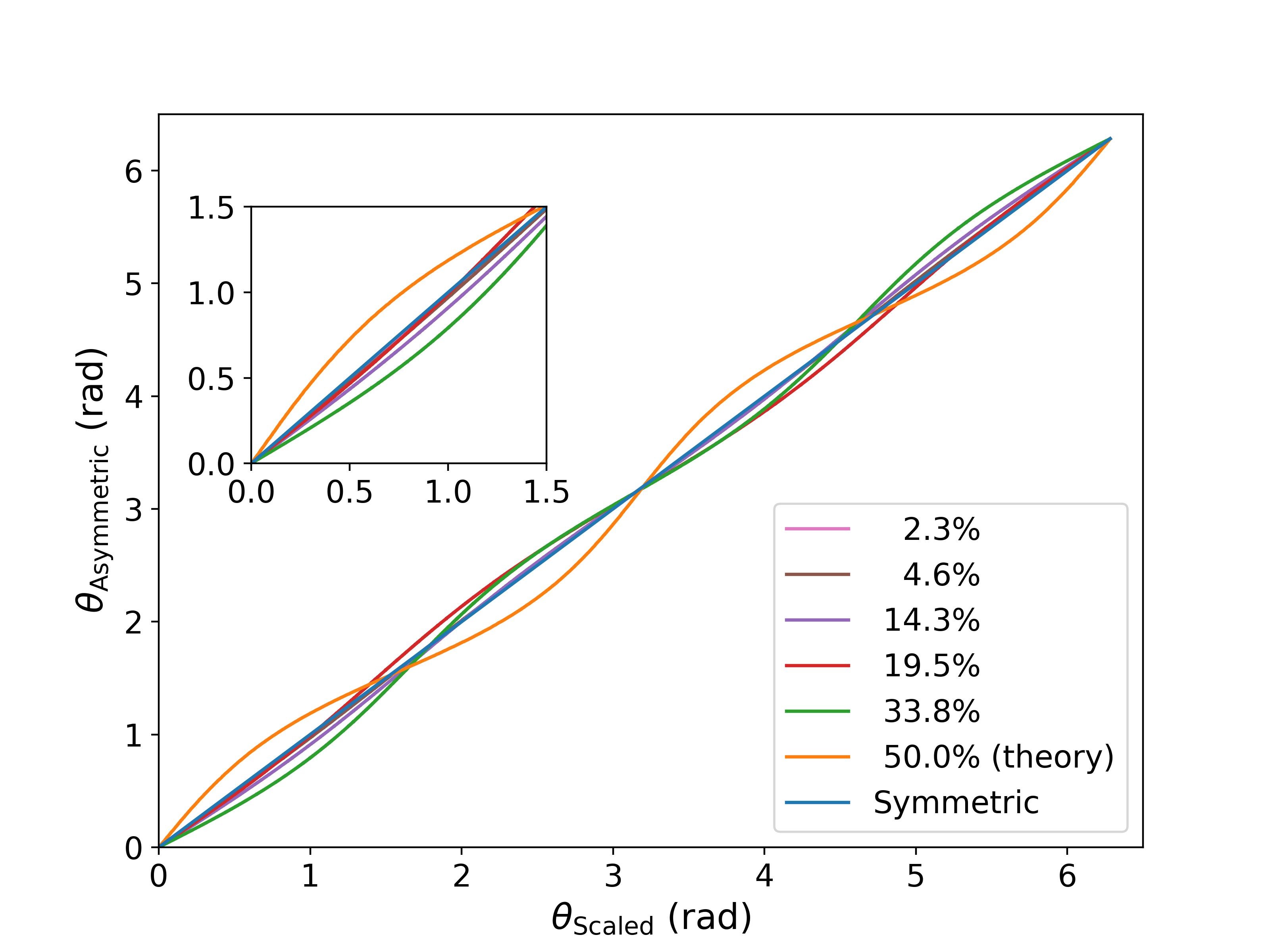}
\caption{Five experimental results (33.8\%---2.3\%) of the deviation in phase between the asymmetric data and scaled data, which we model as being the idealised phase (symmetric detector measurement). At every cyclic value of $\pi/4$ we see the largest deviation. A theoretical asymmetry of 50.2\% is used for demonstrative purposes to show how in extreme cases the phase deviates significantly from the scaled phase.} 
\label{fig:deviationinangles}
\end{figure}

The $X$ and $P$ quadrature outputs were acquired over multiple phase rotations of the reference signal using a fast real-time oscilloscope.  The acquired data $X$ and $P$  is  shown in  Figure \ref{fig:asymmetry} , in orange, for an asymmetry (percentage difference between quadratures) of 14.29\%, see Appendix \ref{appendix: A}. The measured data for asymmetries of 33.77\%,  19.51\%, 4.55\% and 2.25\% were obtained by changing the VOA settings. The reference signal phase, estimated using  Eq \ref{eq:esttheta}, is  plotted in Figure \ref{fig:deviationinangles}, against the scaled phase, which is modeled as the actual/real phase. This is the reference signal phase if it did not experience im-balancing of the heterodyne detection. As described previously, the incorrect phase estimation leads to higher levels of excess noise that ultimately reduces achievable  secure key transmission distance, as shown in Figure \ref{fig:keyrate}. See Appendix \ref{appendix: B} for the secure key rate estimation.

Phase estimation from the symmetrised $X$ and $P$ quadrature  data minimises the excess noise to a negligible level. Symmetrised data is achieved by  estimating  the degree of  the asymmetry and then using it to scale the quadrature values, as discussed in Section \ref{sec:level3}. Asymmetric and symmetric quadrature data are shown in orange and blue, respectively, in Figure  \ref{fig:asymmetry}.

\begin{figure}[!t]
\centering
\includegraphics[width=3.5in]{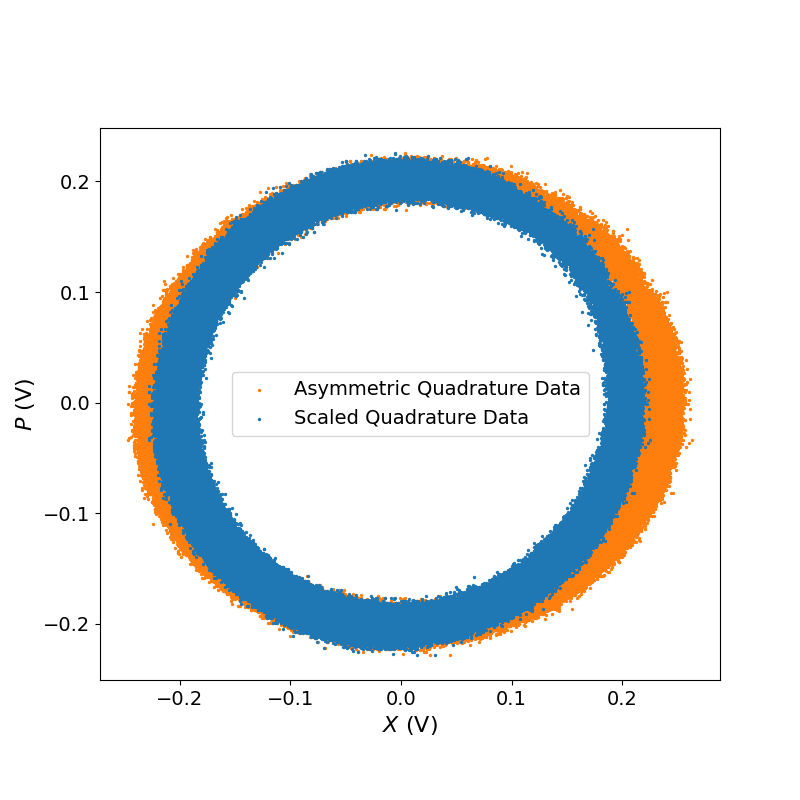}
\caption{A phase space diagram shows a 14.29$\%$ asymmetry in the raw quadrature data, in orange, and scaled quadrature data, in blue.}
\label{fig:asymmetry}
\end{figure}
\begin{figure}[!t]
\centering
\includegraphics[width=3.5in]{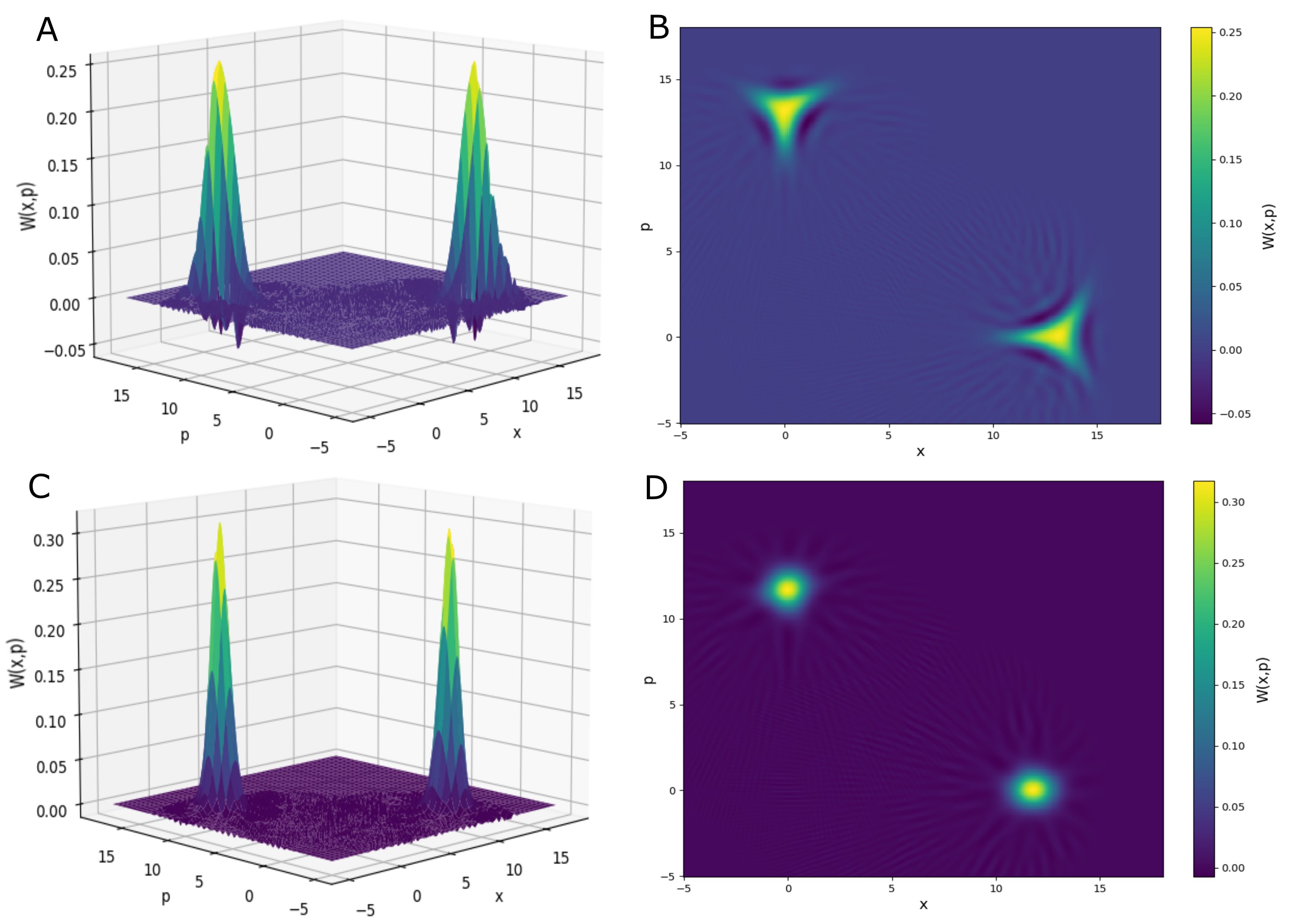}
\caption{The plotted Wigner functions have been calculated using the MLE algorithm \cite{MLEcode2023}, on the two quadrature distributions displayed in Figure \ref{fig:asymmetry}. The 14.29\% detector asymmetry data (orange in Figure 4) produced the Wigner function shown in (a) and (b); the 14.29\% data after scaling (blue in Figure 4) produces the Wigner function shown in (c) and (d). The measured and corresponding simulated state's Wigner function are plotted such that their peaks are centred on (X,0) and (0,P), respectively.}
\label{fig:Wigner Functions}
\end{figure}

To show the influence of  detector asymmetry in quantum tomography, the  quadrature data was used to reconstruct the Wigner function of the  reference signal using Maximum Likelihood Estimation (MLE). For the state reconstruction, the code provided in \cite{MLEcode2023} was used to generate the density matrix, which is the most likely density matrix to produce the recorded quadrature data along with QuTiP's Python library \cite{QuTiP} to create the state's Wigner plot.  The fidelity is calculated with respect to a simulated  ideal coherent state.  As shown in Figure \ref{fig:Wigner Functions} (a) and (b), the Wigner plots are deviated significantly from Gaussian. The fidelity of the state was found to be 0.9205 for an asymmetry of 14.29\%, which improved to 0.9954 after symmetrization. Figure \ref{fig:Wigner Functions} (c) and (d) show the Wigner plots of the reference signal after symmetrization. See Appendix \ref{appendix: C} for fidelity calculations.


\begin{figure}[!t]
\centering
\includegraphics[width=3.5in]{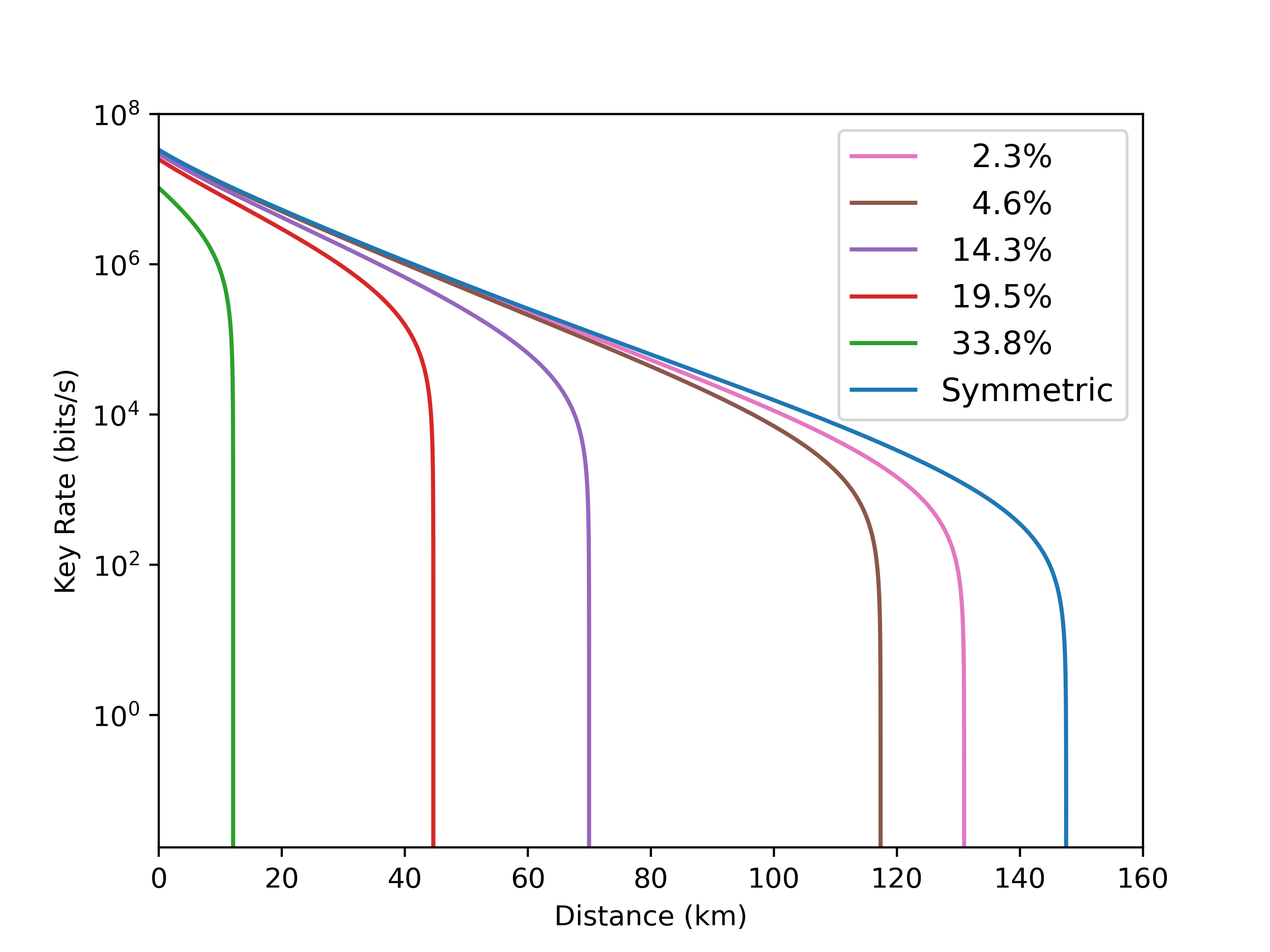}
\caption{Simulated asymptotic key rates for five experimental asymmetric data sets and the symmetrised data (blue). The following parameter values were used to estimate the key rate: $V_A=$10, $\beta=$0.93, $\xi_{line}=$ 0.02 (see equation \ref{eq:eline}), $\eta=$ 0.68, $\nu_{elec}=$ 0.1 and $\alpha=$ 0.2.}
\label{fig:keyrate}
\end{figure}
\section{\label{sec:level6}Conclusion}


We have shown that detection asymmetry degrades performance while using the heterodyne detection scheme for CV-QKD. The asymmetry between the pairs of balanced detectors, assigned to their respective $X$ or $P$ quadrature, can be compensated for by data post-processing.   We propose a quadrature  symmetrisation  method  that reduces the asymmetry-induced phase estimation noise in CV-QKD to improve the transmission distance. We also performed a Wigner state analysis of the quadrature data, where the reconstructed state was far from the ideal Gaussian peak indicative of coherent states, highlighting the detrimental effects of detector asymmetry. The quadrature symmetrisation successfully enables the reconstruction of the state as a Gaussian peak in Wigner representation, improving the state's fidelity with respect to a simulated coherent state.

\section*{Acknowledgements}
JB and RK  acknowledge funding support from the EPSRC Quantum Communications Hub (Grant number EP/T001011/1). AMW would like to acknowledge the support from the EPSRC vacation internship scheme (grant number EP/W524657/1). CJC acknowledges funding from the UK Government’s Department for Science, Innovation \& Technology through the UK national quantum technologies programme.

\section*{Appendices}
\begin{appendices} 

\section{Percentage differences (Asymmetries)}
\label{appendix: A}
We calculate the percentage difference between the input power to each detector in the heterodyne detector set-up, $P_1$ and $P_2$, using
\begin{equation}
\label{eq:percentage difference}
\%_{DIFF}=\frac{|P_1 - P_2|}{\left(\frac{P_1 + P_2}{2}\right)} \times 100.
\end{equation}
As shown in Figure \ref{fig:experimental_set_up_lab}, power input to the $X$ quadrature detector was attenuated compared to that input to the to the $P$ quadrature detector. This mimics non-ideal balancing with respect to each balanced detector.  The power output to the $P$ quadrature detector was kept unchanged at $P_{1}\approx 0.45$ mW whereas the $X$ quadrature detector was changed within a range of $P_2=0.32-0.44$ mW.
\section{Asymptotic key calculation}
\label{appendix: B}
The secret key rate is typically estimated under the collective attack in the reverse reconciliation scenario, therefore the maximum information Eve can receive is bounded by the Holevo information shared between Eve and Bob, $\chi(B;E)$. The key rate can then be given as
\begin{equation}
\label{eq:keyrate1}
r=\frac{B \cdot n}{N} (\beta I(A;B)-\chi(B;E)),
\end{equation}
where $\beta$ is the reconciliation efficiency and $I(A;B)$ is the information shared between Alice and Bob. $B$ is the baud rate and  $n/N$ is the ratio of useful information to the total transmitted information. $I(A;B)$ is given by
\begin{align}
I(A;B)_{\text{het}} &= \frac{1}{2}\log_2 \frac{v_B+1}{v_{B|A}+1} \\
&=\frac{1}{2}\log_2 \frac{(v+\chi_{\text{line}})}{(1+\chi_{\text{line}})}, \\&\quad v=v_A+1\quad\text{and}\quad \chi_{line}=\frac{1}{T}-1+\xi_{\mathrm{ex}}
\end{align}
where $v_A$ is Alice's Gaussian modulation variance, $T$ is the channel transmittance and ``$\text{het}$' refers to the case of heterodyne detection. In the case of detector asymmetry:-
\begin{align}
\xi_{\mathrm{ex}}&=\xi_{\mathrm{RIN}}+\xi_{\text {mod }}+\xi_{\text {quant }}+\xi_{\text {Ram }}+\dots \nonumber\\
& \quad +\{\xi_{\mathrm{drift}}+\xi_{\mathrm{path}}+\xi_{\mathrm{det}}\}\\
&=\xi_{\mathrm{line}}+\xi_{\mathrm{det}}, \label{eq:eline}
\end{align}
where $\xi_{det}$ is the excess noise produced from the asymmetry within the detection scheme. $\xi_{RIN}$, $\xi_{mod}$, $\xi_{quant}$ and $\xi_{Ram}$ are relative intensity, modulation, quantization, and Raman noise, respectively .  $\xi_{det}$= 0.1091, 0.0318, 0.0140, 0.0032 and 0.0016 for asymmetries of 33.77\%, 19.51\%, 14.29\%, 4.55\% and 2.25\%.

\noindent The covariance matrix shared between Alice and Bob is then given by

\begin{align}
\boldsymbol{\gamma}_{\mathrm{AB}}&=\left[\begin{array}{cc}
v\boldsymbol{I} & \sqrt{T \eta} \sqrt{v^2-1} \boldsymbol{Z} \\
\sqrt{T \eta} \sqrt{v^2-1} \boldsymbol{Z} & T \eta(v+\chi_{\text{total}})\boldsymbol{I}
\end{array}\right], \\
&\quad \chi_{total}=\chi_{line} + \frac{\chi_{het}}{T},
\end{align}
\noindent where  $\eta$ is the detection efficiency and  $\chi_{het}=(2-\eta+2v_{elec})/\eta$. From this, a pseudo-matrix can be constructed to determine the Holevo information components later
\begin{equation}
\boldsymbol{\gamma}_{\mathrm{AB}}:=\left[\begin{array}{cc}
a\cdot\boldsymbol{I} & c \cdot\boldsymbol{Z} \\
c\cdot\boldsymbol{Z} & b\cdot\boldsymbol{I}
\end{array}\right].
\label{eq:sudo}
\end{equation}
\noindent The Holevo information can be calculated using the expansion of the characteristic von-Neuman entropy equation:
\begin{align}
\chi(B;E) &= S\left(\rho_{AB}\right) - S\left(\rho_{AB |\text{M}}\right) \\
&\approx g\left(\frac{\lambda_1 - 1}{2}\right) + g\left(\frac{\lambda_2 - 1}{2}\right) \nonumber\\&- g\left(\frac{\lambda_3 - 1}{2}\right) - g\left(\frac{\lambda_4 - 1}{2}\right),
\end{align}
\noindent where $g(x)=(x+1) \log _2(x+1)-x \log _2 x$ is the entropy and 
\begin{align}
\lambda_{1,2} &= \sqrt{\frac{1}{2}\left(A \pm \sqrt{A^2-4B}\right)}, \\
\lambda_{3,4} &= \sqrt{\frac{1}{2}\left(C \pm \sqrt{C^2-4D}\right)}. \\
\end{align}
\noindent It is assumed that Eve can purify the states between Alice and Bob, therefore $S\left(\rho_{AB}\right)$ represents Eve's von Neumann entropy which is equal to the total shared entropy. Similarly, Eve's conditional entropy $S\left(\rho_{E|\text{M}}\right)=S\left(\rho_{AB |\text{M}}\right)$. Using Equation \ref{eq:sudo}, $A, B, C$ and $D$ can be derived from $a, b$ and $c$ as

\begin{align}
A &= v^2(1-2T) + 2T + T^2\left(v+\chi_{\text{line}}\right)^2, \\
B &= T^2\left(1+v\cdot\chi_{\text{line}}\right)^2,\\
C &= (A \cdot \chi_{\text {het }}^2+B+1 \nonumber\\
&+2\cdot\chi_{\text {het }}\left[v \sqrt{B}+T\left(v+\chi_{\text {line }}\right)\right]\nonumber\\ 
&+2 T\left(v^2-1\right))/{T^2\left(v+\chi_{\text {total }}\right)^2},\\
D &= \left(v+\chi_{\text {het }} \sqrt{B}\right)^2/{T^2\left(v+\chi_{\text {total }}\right)^2}.
\end{align}

\section{Quantum Fidelity calculation}
\label{appendix: C}

Let \( \rho \) be the density matrix of the measured state and \( \sigma \) be the density matrix of the simulated state as determined by the MLE algorithm, then the fidelity \( F(\rho, \sigma) \) is given by:

\begin{equation}
F(\rho, \sigma) = \left( \text{Tr} \left[ \sqrt{ \sqrt{\rho} \sigma \sqrt{\rho} } \right] \right)^2
\end{equation}

Where  \(\text{Tr} \)   represents the trace of the matrix and with the requirements that \(\sqrt{\rho} \)  is the matrix square root of \( \rho \), such that \( (\sqrt{\rho})^2 = \rho \) and that $\rho$ and $\sigma$ have the same dimensions.

\subsection*{QuTiP.fidelity}
    As only the trace of the intermediate product is required, for optimisation, the matrix  is diagonalised, enabling its eigenvalues to be calculated. Denoting these eigenvalues as \( \lambda_i \), QuTiP calculates the fidelity as follows:
    
\begin{equation}
    \text{Fidelity} = \sum_{i} \sqrt{\lambda_i}
\end{equation}

    Note that small negative values may be present and are truncated to prevent NaN (not a number) values - even for positive semi-definite matrices, small negative eigenvalues can be reported.

\end{appendices}

\bibliographystyle{IEEEtran} 
\bibliography{bib} 

\begin{IEEEbiographynophoto}{Jennifer Bartlett}
is a PhD student at the University of York (UK). Her research interests include practical security in fibre-based CV-QKD systems. Contributions include concept, experimental design, implementation, post-processing analysis and drafting the manuscript.
\end{IEEEbiographynophoto}
\begin{IEEEbiographynophoto}{Alfie Myers Wilson}
is a MSc student at the University of York (UK). His research interests include quantum technologies. Contributions include Wigner function analysis, fidelity calculation and editing parts of the manuscript.
\end{IEEEbiographynophoto}


\begin{IEEEbiographynophoto}{Christopher Chunnilall} is a Principal Scientist at the National Physical Laboratory (UK). His research interests are in single-/few-photon metrology and quantum communications. Contributions include investigating detector asymmetry and reviewing and editing the manuscript.
\end{IEEEbiographynophoto}

\begin{IEEEbiographynophoto}{Rupesh Kumar} 
is a Lecturer in Experimental Quantum Communications at the University of York (UK). Contributions include identifying detector asymmetry and conceptualising the impact in CVQKD, leading the project, supervising the research work and reviewing and editing the manuscript. 
\end{IEEEbiographynophoto}
\newpage

\vfill

\end{document}